\documentclass[twocolumn,showpacs,preprintnumbers,amsmath,amssymb,prb,superscriptaddress]{revtex4}
\usepackage{graphicx}% Include figure files
\usepackage{dcolumn}% Align table columns on decimal point
\usepackage{bm}% bold math

\begin{document}
\title{Single-particle Excitation Spectra of C$_{60}$ Molecules and Monolayers}
\author{Fei Lin}
 \affiliation{Department of Physics, University of Illinois at Urbana-Champaign, Urbana, Illinois 61801, USA}
\author{Erik S. S$\o$rensen}
 \affiliation{Department of Physics and Astronomy, McMaster University, Hamilton, Ontario,
 Canada L8S 4M1}
\author{Catherine Kallin}
 \affiliation{Department of Physics and Astronomy, McMaster University, Hamilton, Ontario,
 Canada L8S 4M1}
\author{A. John Berlinsky}
 \affiliation{Department of Physics and Astronomy, McMaster University, Hamilton, Ontario,
 Canada L8S 4M1}
 \date{\today}

\begin{abstract}
In this paper we present calculations of single-particle excitation spectra of
neutral and three-electron-doped Hubbard C$_{60}$ molecules and
monolayers from large-scale quantum Monte Carlo simulations and
cluster perturbation theory. By a comparison to experimental
photoemission, inverse photoemission, and angle-resolved
photoemission data, we estimate the intermolecular 
hopping integrals and the C$_{60}$ molecular
orientation angle, finding agreement with recent X-ray
photoelectron diffraction (XPD) experiments. Our results 
demonstrate that a simple effective Hubbard model, with intermediate
coupling, $U=4t$,  provides a 
reasonable basis for modeling the properties of C$_{60}$ 
compounds.
\end{abstract}


\maketitle

\section{Introduction} 
C$_{60}$ compounds occupy a unique place within the pantheon of 
high-temperature superconductors. \cite{hebard91, rosseinsky91, 
tanigaki91, holczer91, kelty91, fleming91} One would not normally expect such 
high transition temperatures (30-40K) to arise from the conventional 
phonon mechanism of superconductivity. However, the unusual molecular 
structure of these compounds results in higher frequency phonons and 
stronger couplings than are possible in typical elemental compounds and 
alloys. \cite{schluter92} In a similar way, the 
effects of electron-electron interactions on these molecules are more 
complex and, it has been argued, \cite{kivelson91, baskaran91} 
could also provide the mechanism for attractive pairing. 
There is general agreement that the attractive mechanism is 
intramolecular, i.e., it arises from intramolecular vibrations or from Coulomb 
correlations on a molecule or, perhaps most likely, from a combination of 
the two effects. \cite{han03}  This intramolecular attraction, together with the narrow bands and 
correspondingly large density of states (DOS) for intermolecular hopping, is 
believed to account for the observed high transition temperatures. The two 
contrasting theoretical approaches, which emphasize either phonons or 
Coulomb interactions, can be roughly characterized as an LDA electronic 
structure coupled to Jahn-Teller phonons, compared to a Hubbard model of 
the C$_{60}$ molecule with tight-binding intermolecular hopping.
\begin{figure}
  \centering
  \begin{tabular}{c}
  \resizebox{70mm}{!}{\includegraphics{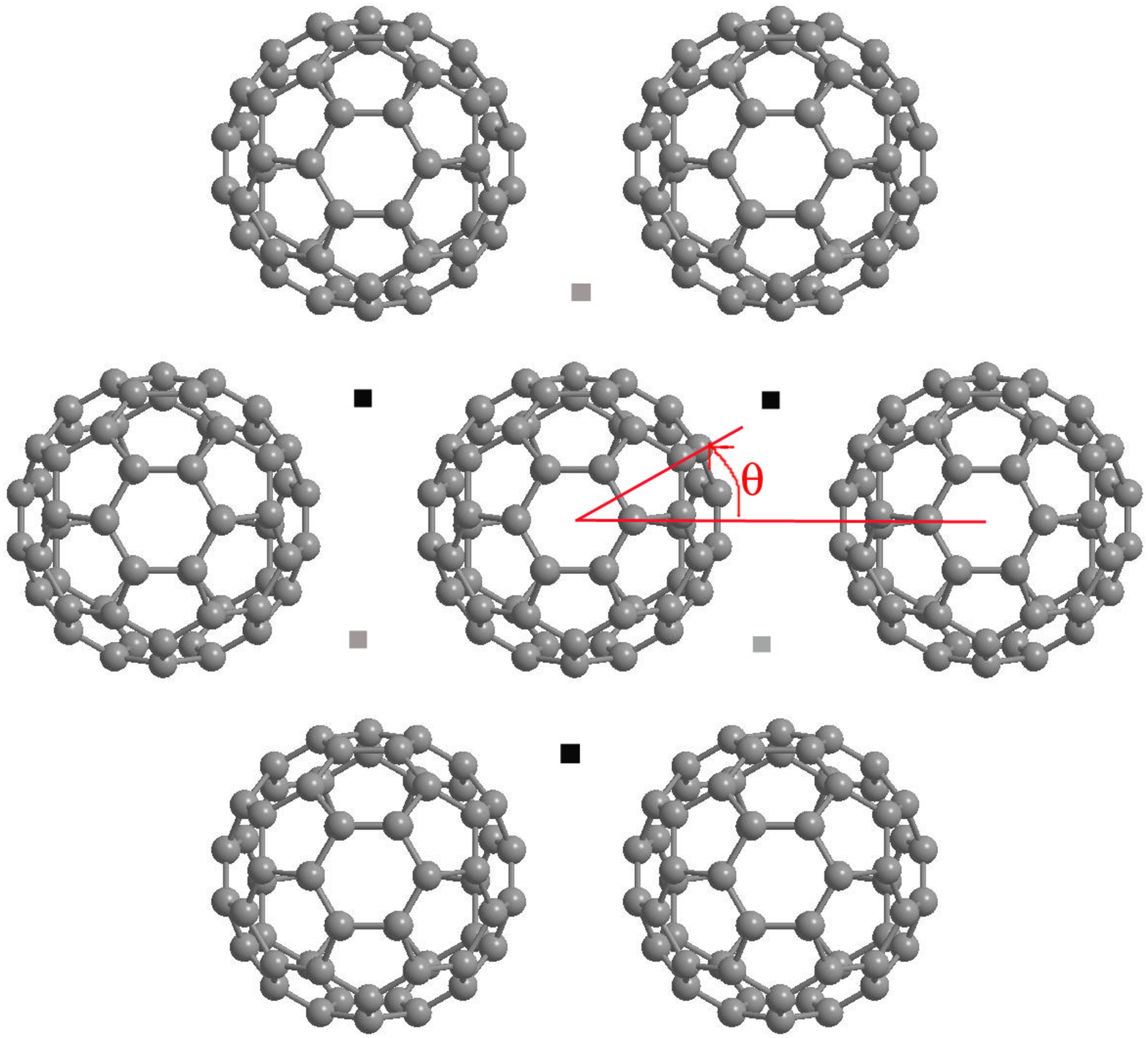}} \\
  \resizebox{20mm}{!}{\includegraphics{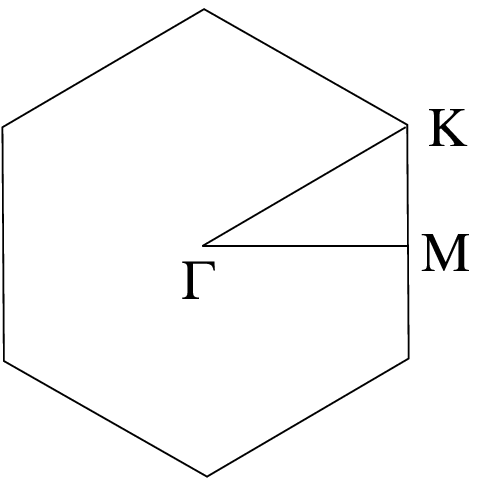}} \\
  \end{tabular}
  \caption{(Color online) 2D hexagonal monolayer of identically oriented C$_{60}$ molecules (lattice 
    constant $a=10.02\AA$ for the superlattice \cite{shen03}) and its
    corresponding first Brillouin zone (BZ). The definition of the molecular rotation 
    angle $\theta$ is the same as in the caption of Fig. 4 of Ref.\ \onlinecite{shen03}, 
    which is defined by two lines, both projected onto the plane of the layer, 
    one from a molecular center to a NN molecular center,
    as shown, and the other from a
    molecular center through a pentagon face center. $\theta=0^{o}$ when these two lines 
    coincide. Counterclockwise rotation of the molecule corresponds 
    to a positive rotation angle. For K$_3$C$_{60}$, the black squares 
    represent two K$^{+}$ ions, one above the 
    other, while gray ones denote a single K$^{+}$ ion. \cite{shen03}}
  \label{c60layer}
\end{figure}

Recently Yang \textit{et al.} \cite{shen03} reported angle-resolved photoemission (ARPES) and LDA 
calculations of the band structure of two-dimensional (2D) hexagonal 
monolayers of C$_{60}$ and K$_3$C$_{60}$ on the surface of silver, and Wehrli \textit{et al.} 
\cite{wehrli04} have proposed an electron-phonon mechanism to explain the DOS observed in 
Yang \textit{et al.}'s study as well as spectra from earlier experiments. 
\cite{chen91, benning93, hesper00} Since their calculations do not capture the distinctive correlation 
effects of the Hubbard model within a molecule, it is of interest to examine how the DOS and other 
properties are affected by such correlations.

In this paper, we calculate the photoemission spectra (PES) and
inverse photoemission spectra (IPES) of 2D hexagonal C$_{60}$
monolayers including the strong, atomic, on-site electron-electron Coulomb
interaction $U$ within a single molecule. The calculation is
performed by cluster perturbation theory (CPT) \cite{senechal00,
senechal02} using quantum Monte Carlo (QMC) data. \cite{lin06} This method, which we call 
QMCPT, is ideally suited to calculating the electronic properties of solids composed of 
complex units such as fullerene molecules. To distinguish, we shall refer to the standard
CPT approach \cite{senechal00, senechal02} as EDCPT since it uses exact diagonalization (ED) results.
By comparing the calculated single-particle excitation spectra, for intermediate interaction 
strength, $U=4t$, with experimental data, \cite{lof92, shen03} we can estimate the intermolecular 
hopping integrals $t^{'}$ and the molecular orientation angle, $\theta$, which are useful for an 
effective Hubbard model Hamiltonian description of C$_{60}$ compounds. From the best fit parameters,
we find a rotation angle of $\theta=64^{o}$ (as defined in Fig.\ \ref{c60layer}, and assuming that 
the C$_{60}$ molecules are all oriented identically) which is consistent with experimental 
measurements. \cite{shen03, tamai05}

\section{Model} 
The Hamiltonian for the 2D hexagonal C$_{60}$ superlattice is given by
\begin{eqnarray}
H&=&H_0+V,\\
H_0&=&\sum_I H_0^I,\label{modelhamiltonian}
\end{eqnarray}
where
\begin{equation}
H_0^I=-\sum_{\langle Ii,Ij\rangle \sigma}
t_{ij}(c^{\dagger}_{i\sigma}c_{j\sigma}+h.c.)+U\sum_{i\in I}
n_{i\uparrow}n_{i\downarrow} \label{cterm}
\end{equation}
is the Hubbard model on the $I$th C$_{60}$ molecule, and
\begin{equation}
V=-\sum_{\langle Ii,Jj \rangle\sigma}
(t^{'}\alpha_{Ii,Jj}+t^{\hbox{ind}}_{Ii,Jj})
(c^{\dagger}_{Ii\sigma}c_{Jj\sigma}+h.c.)
\label{Vterm}
\end{equation}
is the hopping Hamiltonian between a pair of carbon (C) atoms $i$ and $j$
on two nearest neighbor (NN) C$_{60}$ molecules $I$ and $J$,
respectively. The operator notations are standard for the
usual Hubbard model. Inside the molecule $U$ is the on-site
Coulomb interaction energy for two electrons on the same C atom, and
$t_{ij}$ is the hopping integral between two NN carbon atoms. A {\it molecular} 
$U$, parameterizing the interaction energy for two electrons on the same 
C$_{60}$ {\it molecule}, and hence distinct from our atomic $U$, is also discussed 
in the literature. See for instance Ref.\ \onlinecite{lof92}. For the
C$_{60}$ molecule, there are two kinds of bonds between NN sites:
$t_{ij}=t$ for single bonds ($1.46\AA$), and $t_{ij}=1.2t$ for
double bonds ($1.40\AA$). We set $t=2.72$eV according to Ref.\
\onlinecite{satpathy86}. 
The value of the on-site Coulomb energy, U, is taken to be $U=4t$,
an intermediate coupling value for which the sign problem for the
single-molecule QMC is manageable. \cite{lin05}

The shortest distance between two C$_{60}$ molecules is about
$3.0\AA$. The relative sizes of the direct hopping integrals $\alpha_{Ii,Jj}$ between two
C atoms on two NN molecules are given by \cite{satpathy92}
\begin{equation}
\alpha_{Ii,Jj}=[V_{\sigma}(d)-V_{\pi}(d)](\hat{\textbf{R}}_{Ii}\cdot\hat{\textbf{d}})
(\hat{\textbf{R}}_{Jj}\cdot\hat{\textbf{d}})+V_{\pi}(d)(\hat{\textbf{R}}_{Ii}\cdot
\hat{\textbf{R}}_{Jj}),
\end{equation}
where $\hat{\textbf{R}}$ is a unit vector along radial direction,
$\hat{\textbf{d}}=\textbf{d}/d$ is a unit vector pointing from
atom $Ii$ to $Jj$, and
\begin{equation}
V_{\sigma}(d)=-4V_{\pi}(d)=\frac{d}{d_0}\exp[-(d-d_0)/L],
\end{equation}
with $L=0.505\AA$, and $d_0=3.0\AA$. \cite{satpathy92} We leave the
overall prefactor $t^{'}$
as a parameter to be determined later. 

We see from Fig.\ \ref{c60layer} that for K$_3$C$_{60}$ the C$_{60}$ molecules are separated by 
K$^{+}$ ions, either two ions, one above the other or single ions. The indirect
hopping between C$_{60}$ molecules via K$^{+}$ ions is thus of comparable importance 
to that of the direct C-C hopping integrals.  
Expressions for the indirect hopping between 
C atoms via the K$^{+}$ ions are given by \cite{gunnarsson98}
\begin{equation}
t^{\hbox{ind}}_{Ii,Jj} =\sum_{\gamma}\frac{t_{Ii,\gamma}t_{Jj,\gamma}}{\epsilon_C-\epsilon_K},
\end{equation}
where $\epsilon_C-\epsilon_K=-4$eV is the energy difference between a C atom and 
K$^{+}$ ion, $\gamma=1,\cdots,9$ is K$^{+}$ index, and the C-K hopping integral 
is given by
\begin{equation}
t_{Ii,\gamma}=1.84D\frac{\hbar^2}{md^2}e^{-3(d-R_{\hbox{min}})},
\end{equation}
where $d$ is C-K separation, $m$ is electron mass, $R_{\hbox{min}}$ is the 
shortest C-K separation for a given K$^{+}$ and C$_{60}$ molecule, and $D=0.47$. 
\cite{gunnarsson98} We take the indirect hopping matrix elements to be fixed while
the direct hopping matrix elements scale with the fitting parameter $t^{'}$.  For the value
of $t^{'}$ used to fit the ARPES data, we find that the largest direct hopping matrix 
elements are substantially larger than the indirect hopping matrix elements.
\begin{figure}
  \begin{tabular}{c}
  \resizebox{80mm}{!}{\includegraphics{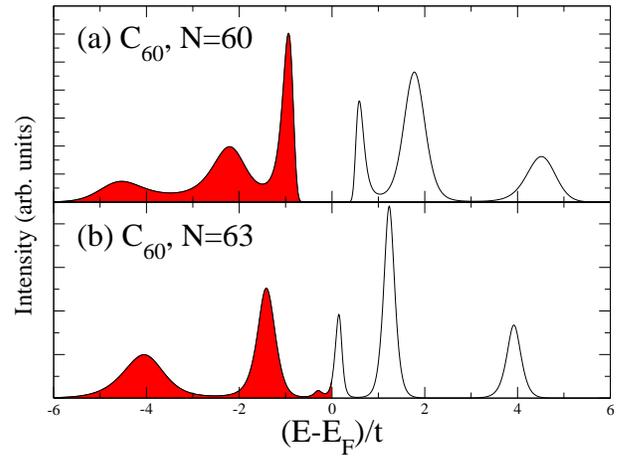}}
  \end{tabular}
  \caption{(Color online) DOS of (a) C$_{60}$ and (b) C$_{60}^{3-}$ single molecules from QMC and MEM 
  for $U=4t$ and $\beta t=10$. A total of 465 and 620 bins of space-time Green's functions 
  are collected for $C_{60}$ and $C_{60}^{3-}$, respectively, for MEM analysis. Each bin is
  an average over 100 space-time Green's functions, which are collected after
  each QMC sweep over the whole space-time lattice.}
  \label{c60dos}
\end{figure}

We note that our model Hamiltonian, given by Eqs.\ (\ref{modelhamiltonian})-(\ref{Vterm}), 
neglects both the long-range Coulomb interactions and the effect of the
silver substrate, both of which would be difficult to include
in QMC calculations. One would expect the long-range Coloumb interaction
to increase the effective molecular $U$. On the other hand, the silver
substrate, as well as the other molecules, will screen this interaction
and, consequently, these two effects may partially cancel. In any case,
we will see below that the simple Hubbard model and parameters employed
here yield results in generally good agreement with experiment.

\section{Results}
\subsection{Single Molecule DOS}
We begin by considering results for a single C$_{60}$ molecule. 
QMC simulations were used to calculate the imaginary-time
Green's functions for C$_{60}$ and C$_{60}^{3-}$ molecules,
based on the single-molecule Hamiltonian, $H_0$ and standard QMC 
\cite{hirsch85, hirsch88, white89} for given values of $t$ and $U$.  
The QMC step for the single molecule is by far the most
time-consuming part of the calculation.
The maximum entropy method (MEM) was then used to analytically
continue the imaginary-time Green's functions to real frequency,
\cite{jarrell96} which gives the single molecule DOS shown in 
Fig.\ \ref{c60dos} for $T=0.1t$. From the figure, we see
that neutral C$_{60}$ is an insulator with an energy gap about
$1.05t$ or $2.86$eV. The presence of this gap is not surprising 
since neutral C$_{60}$ has a filled shell with a sizable excitation gap 
even for $U=0$. In fact the gap obtained by extended Huckel calculations
for the neutral model is also about $1t$, although the correlations 
for that case are very different than for $U=4t$.  For the C$_{60}^{3-}$ molecule,
the Fermi energy lies in a region of non-zero density of states.
\begin{figure}
  \begin{tabular}{c}
  \resizebox{80mm}{!}{\includegraphics{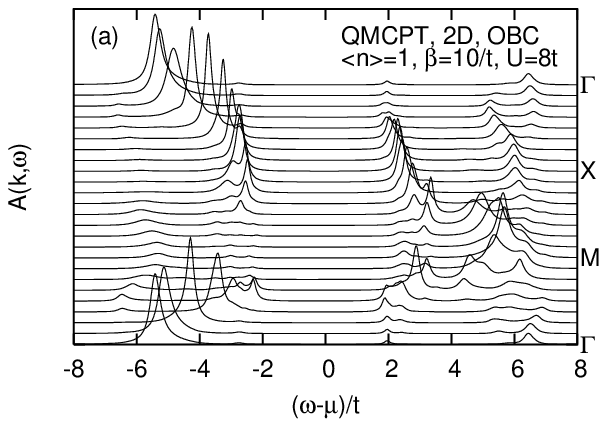}} \\
  \resizebox{80mm}{!}{\includegraphics{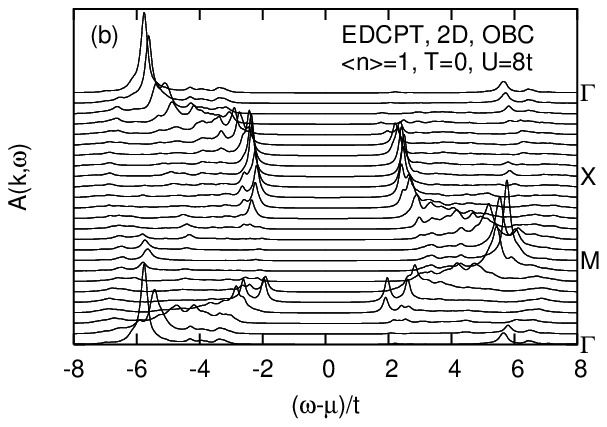}}\\
  \end{tabular}
  \caption{Single particle spectral functions of a 2D Hubbard model
  from (a) QMCPT and (b) EDCPT with open boundary conditions.
  The cluster is of dimension $3\times 4$. The two methods predict very similar
  single-particle excitation energies and energy gaps. Note that special \textbf{k}
  points $\Gamma, M, X$ are for a 2D square lattice, different from those in 
  Fig.\ \ref{c60layer}.}
  \label{2du8}
\end{figure}

\subsection{QMCPT for 2D Square Lattice}
The remaining perturbative part of the QMCPT calculation is, by comparison, very economical.
It uses the single molecule Green's functions calculated above to construct any 
superlattice Green's functions, incorporating the effect of the intermolecular 
hopping Hamiltonian, $V$, and different molecular orientations 
with negligible additional computational effort. This method has been applied to 
the simpler case of the 2D Hubbard model on a square lattice, which is metallic 
for $U=0$ but which develops an insulating Mott-Hubbard gap for sufficiently large $U$.
The ``cluster'' in this case is a small unit of the lattice which can be treated
either by QMC or by exact diagonalization. Results for $U=8t$, obtained by QMCPT 
and by EDCPT, based on a $3\times 4$ cluster of sites, \cite{lin06} are shown in Fig.\ \ref{2du8}.

\subsection{Monolayer DOS}
\begin{figure}
  \begin{tabular}{c}
  \resizebox{90mm}{!}{\includegraphics{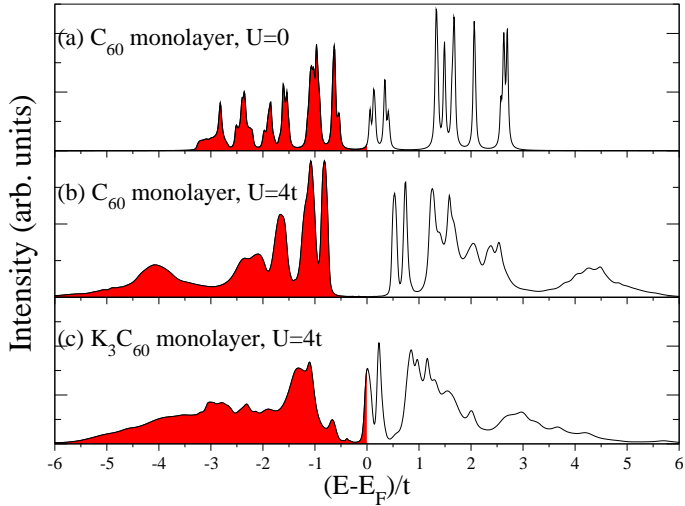}}
  \end{tabular}
  \caption{(Color online) DOS of hexagonal (a) C$_{60}$ monolayer with $U=0$, 
   (b) C$_{60}$ monolayer with $U=4t$, and
   (c) K$_3$C$_{60}$ monolayer with $U=4t$ from QMCPT calculations for $t^{'}=-0.3t$ and $T=0.1t$.
   Shaded areas are occupied by electrons.}
  \label{c60hex}
\end{figure}

Next we consider the case of C$_{60}$ monolayers. For these calculations we use values of
$t^{'}$ and the orientation angle $\theta$ which are obtained by fitting photoemission
and diffraction data as will be discussed below. We show the DOS for monolayers of
both C$_{60}$ and K$_3$C$_{60}$, in panels (b) and (c) of Fig.\ \ref{c60hex}. 
Also shown for comparison, in panel (a) of Fig.\ \ref{c60hex}, is the DOS for the same values
of $t^{'}$ and $\theta$, but for $U=0$.  The lattice of neutral C$_{60}$ molecules is very different
from the 2D square lattice of atomic sites, for which spectra are shown in Fig.\ \ref{2du8}, because the 
C$_{60}$ molecule has an excitation gap as was shown in the top panel of Fig.\ \ref{c60dos}.  
For $U=0$ [(a) panel of Fig.\ \ref{c60hex}], the C$_{60}$ monolayer 
is a band insulator with a gap of about $0.5t$ which is about half the value for the single-molecule case 
because the bands above and below the gap are brodened by temperature effects and the intermolecular 
Hamiltonian $V$, thus reducing the gap. We see, from the (b) panel where $U=4t$, that the effect of $U$
is to enlarge the gap back to a value ($\sim 2.8$eV) slightly greater than $1t$ due to short-range, 
Mott-Hubbard corelations. This gap value is comparable to the value of $2.3$eV measured by Lof 
\textit{et al.} in Ref.\ \onlinecite{lof92}. Our calculated gap value may be somewhat larger than 
the experimental gap due to our neglect of the silver substrate, which may reduce the energy gap. 
\cite{hesper97} By contrast, for the case of K$_3$C$_{60}$ [(c) panel] the monolayer remains 
metallic with its Fermi energy lying within a narrow conduction band (Fig.\ \ref{c60hex}).

\begin{figure}
  \begin{tabular}{c}
  \resizebox{80mm}{!}{\includegraphics{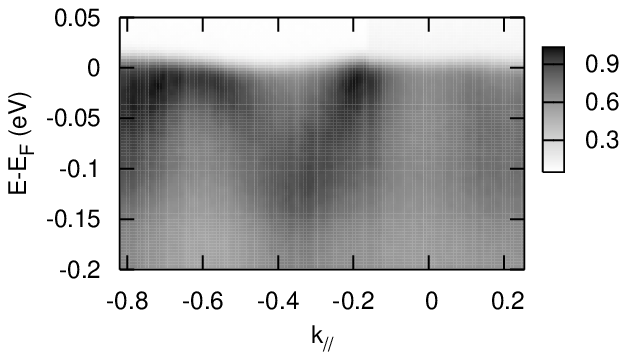}} \\
  \resizebox{80mm}{!}{\includegraphics{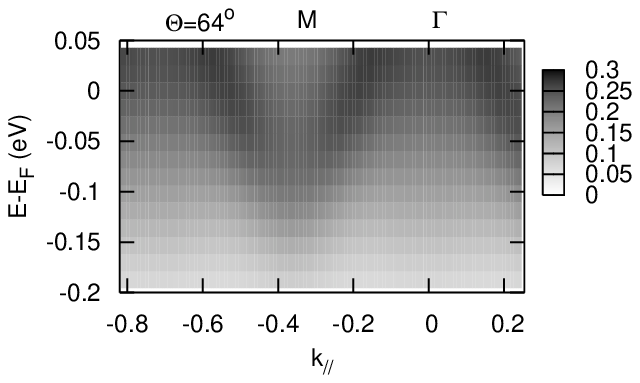}}\\
  \end{tabular}
  \caption{Determination of NN molecular hopping integral $t^{'}$ and molecular orientation 
  angle $\theta$ by comparing the experimental ARPES result \cite{shen03} (upper panel) with
  the theoretical one (lower panel), which has $t^{'}=-0.3t$ and molecular rotation angle 
  $\theta=64^{o}$. The energy unit has been converted to eV using $t=2.72$eV. The momentum 
  unit is $\AA^{-1}$.}
  \label{tpfit}
\end{figure}

To determine the parameter $t^{'}$ and orientation angle $\theta$, we vary 
$t^{'}$ and $\theta$ values in the QMCPT calculations. The
resulting energy-momentum dispersion curves (along the $\Gamma M$
direction) for the hexagonal monolayer of K$_3$$C_{60}$ are then 
compared with the experimental one from ARPES. \cite{shen03} We find that the
best fit is obtained when $t^{'}=-0.3t$ with a rotation angle
$\theta=64^{o}\pm 4^{o}$. From Fig.\ \ref{tpfit}, we see
that both the experimental and theoretical band widths are about
100meV. However the theoretical band dispersion in the $k$ range
$[-0.8, -0.6]\AA^{-1}$ is rather weak around the Fermi energy, while in
the experimental dispersion figure, the band dispersion in that
momentum range is much more obvious. We think that this results from two causes:
First MEM has difficulty reconstructing the weak features of the band
dispersion curve, and, second, the experimental scanning direction is not
exactly along $\Gamma M$. \cite{shen03} The above determined
rotation angle $\theta=64^{o}\pm 4^{o}$ is  in good agreement with XPD structural data. \cite{tamai05}
\begin{figure}
 \begin{tabular}{c}
 \resizebox{80mm}{!}{\includegraphics{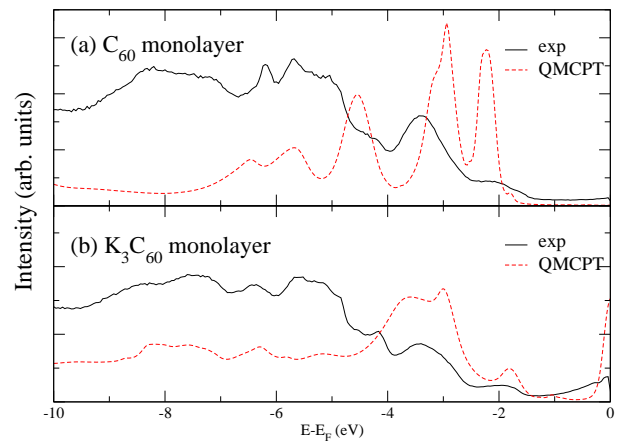}}
 \end{tabular}
 \caption{(Color online) Comparison of DOS from QMCPT calculations and experimental PES data \cite{shen03}
 for C$_{60}$ (upper panel) and K$_3$C$_{60}$ (lower panel) hexagonal monolayer.
 The QMCPT energy scale has been converted to electron volts by setting
 $t=2.72$eV.}
 \label{c60hexpes}
\end{figure}

\begin{figure*}
  \begin{tabular}{cc}
  \resizebox{80mm}{!}{\includegraphics{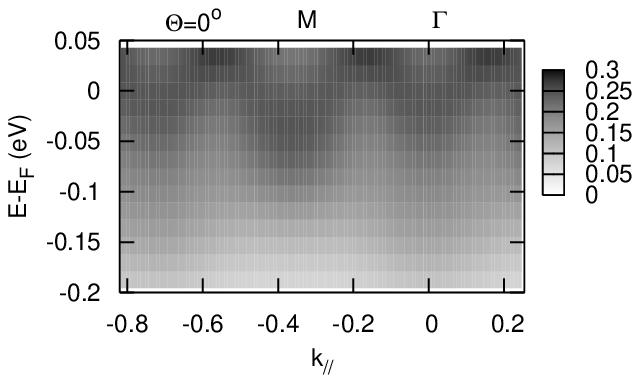}}&
  \resizebox{80mm}{!}{\includegraphics{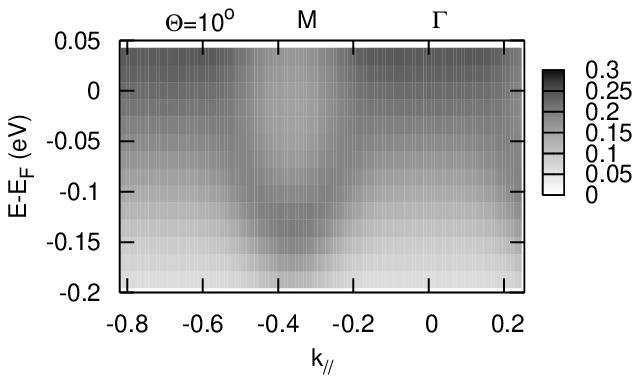}}\\
  \resizebox{80mm}{!}{\includegraphics{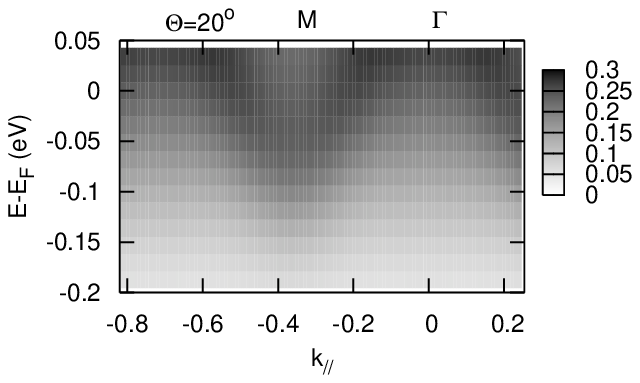}}&
  \resizebox{80mm}{!}{\includegraphics{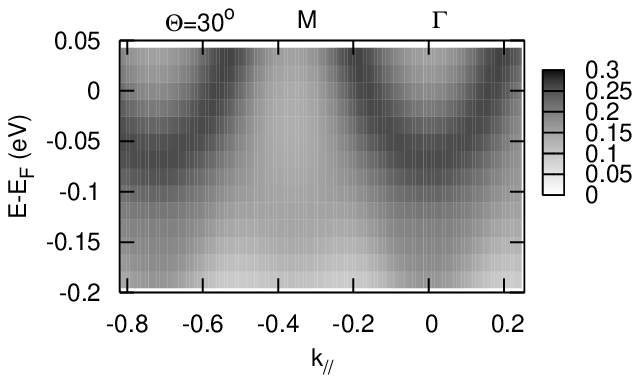}}\\
  \resizebox{80mm}{!}{\includegraphics{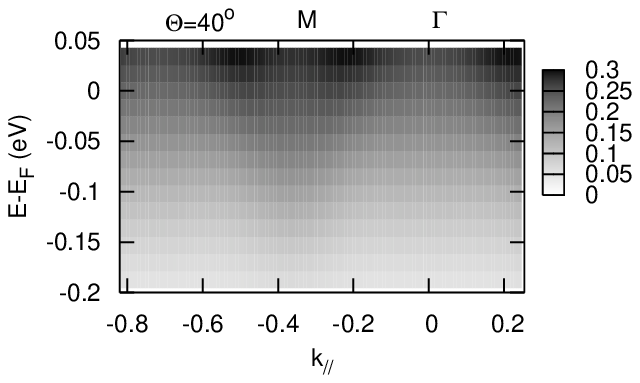}}&
  \resizebox{80mm}{!}{\includegraphics{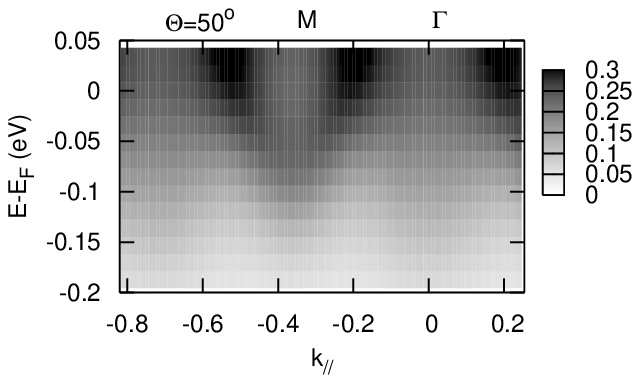}}\\
  \resizebox{80mm}{!}{\includegraphics{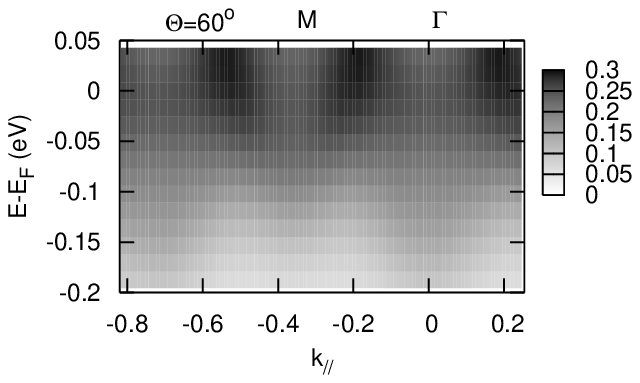}}&
  \resizebox{80mm}{!}{\includegraphics{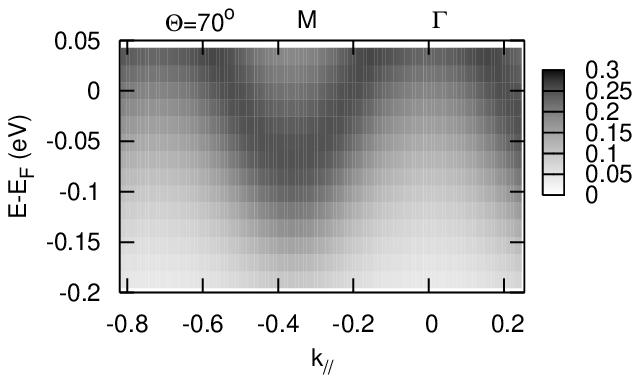}}\\
  \end{tabular}
  \caption{Band dispersions along $\Gamma M$ direction
    with different rotation angles $\theta$ for a 2D hexagonal K$_3$C$_{60}$
    superlattice from QMCPT calculations. The other parameters are fixed: $U=4t$, 
    $t^{'}=-0.3t$, and $T=0.1t$. The momentum unit is $\AA^{-1}$.}
  \label{c60angvary}
\end{figure*}
The densities of states in the lower two panels of Fig.\ \ref{c60hex} are directly
comparable to the PES data (Fig.\ 1E in
Ref.\ \onlinecite{shen03}) of Yang \textit{et al.}, we draw only the photoemission (PE)
part of the calculated DOS together with the experimental PES data in Fig.\ \ref{c60hexpes}.
We see that the QMCPT results agree reasonably well with the PES data close to the Fermi energy,
with approximately the same energy gaps
for the C$_{60}$ monolayer and similar peaks at the Fermi energy for the K$_3$C$_{60}$ monolayer.
In addition, we find small peaks at around $-2$eV and valleys at
around $-4$eV for both theoretical and experimental PES curves. Thus, although we do not claim that
$U=4t$ corresponds to a fit to the data, this value of $U$ is consistent with the PES data.  Smaller 
or larger values of $U$ would lead to smaller or larger values of the gap for neutral C$_{60}$ monolayers. 
Unfortunately, to go beyond this by varying the value of $U$ would, at present, require a prohibitively 
large amount of computer time.

We comment that the above-determined value of $t^{'}$ is consistent with earlier TB and LDA calculations
\cite{satpathy92}, where Satpathy \textit{et al.} found good agreement between TB and LDA band
structure calculations for a fcc C$_{60}$ lattice. The magnitudes of their TB hopping integrals 
between two nearest C atoms on two NN C$_{60}$ molecules are, depending on the relative orientations 
of NN C$_{60}$ molecules, in the range of $(0.50\textrm{eV},0.82\textrm{eV})$. From our calculation, 
the maximum C-C hopping integral between two NN C$_{60}$ molecules is about $0.51$eV (calculated 
from $t^{'}\alpha_{Ii,Jj}$). Therefore, our estimated TB hopping integrals agree with 
Satpathy \textit{et al.}'s calculations.
				
In our formalism, it is easy to explore 
the variation of the monolayer band structure with different C$_{60}$  rotation angles $\theta$. 
In fact it was necessary to do just this to determine the orientation which best fit the ARPES 
band structure. Fixing $t^{'}=-0.3t$, we 
carry out QMCPT for C$_{60}$ molecular rotation angles $\theta=0, 10^{o}, \cdots, 70^{o}$ for a 
K$_3$C$_{60}$ monolayer. The band dispersions along the $\Gamma M$ direction are shown in 
Fig.\ \ref{c60angvary}. We see that the band structure is quite sensitive to the molecular 
rotation angle $\theta$. For example, at $\theta=30^{o}$, the band is inverted with a band minimum at 
$\Gamma$ and a maximum at $M$. At $\theta=40^{o}$, the band becomes almost 
insulating. These are quite interesting, and might be verified by experiments if a way can be 
found to control the molecular orientation.

\section{Conclusion} 
To summarize we have calculated theoretical single-particle 
excitation spectra for single C$_{60}$
and C$_{60}^{3-}$ molecules as well as for hexagonal monolayers of C$_{60}$ and K$_3$C$_{60}$. The
calculations were carried out by the QMCPT technique, \cite{senechal00, senechal02, lin06} 
using a Hubbard model for each molecule, and simple electron 
hopping between molecules of the 2D C$_{60}$ superlattice.  This new computational method is 
particularly well-suited to composite systems of fullerenes.
Using a plausible value,
$U=4t$, for the on-site Coulomb interaction $U$, we fit the effective NN molecular hopping
amplitude, $t^{'}$, and the C$_{60}$ molecular rotation angle $\theta$ to ARPES and XPD data. 
We find that the electronic spectra are extremely sensitive to the molecular orientation
angle $\theta$.
With these fitted parameters, we 
find good agreement with photoemission data for the electronic density of states. 
The C-C hopping integrals between NN molecules obtained from the fits are consistent with
earlier TB and LDA calculations. \cite{satpathy92} 

The significance of the present calculations is to show that the electronic spectral data obtained so 
far are compatible with the relatively simple model of electronic correlations contained 
in the intramolecular Hubbard model with simple interball hopping, in much the same way as they 
have also been shown by others \cite{shen03,wehrli04} to be compatible with LDA/phonon 
calculations. Further measurements and calculations will be required in order to refine our 
understanding of the dominant correlations in these systems.

\begin{acknowledgments}
This project was supported by the Natural Sciences and Engineering
Research Council (NSERC) of Canada, the Canadian Institute for
Advanced Research (CIAR), and the Canadian Foundation for
Innovation (CFI). FL was supported by US Department of Energy 
under award number DE-FG52-06NA26170. AJB and CK gratefully acknowledge 
the hospitality and support of the Stanford Institute for Theoretical Physics where part
of this work was carried out. All the calculations were performed at
SHARCNET supercomputing facilities. We thank Z.X. Shen and W.L.
Yang for helpful discussions and for sending us the experimental data
shown in Fig.\ \ref{tpfit} and \ref{c60hexpes}. We also thank G.A. Sawatzky 
for helpful discussions. 
\end{acknowledgments}

\bibliography{c60specbib}

\end{document}